\documentclass{ws-procs9x6-cpt19}

\def\al{\alpha}
\def\be{\beta}

\def\de{\delta}

\def\ka{\kappa}
\def\la{\lambda}

\def\si{\sigma}

\def\ph{\phi}

\def\fr#1#2{{{#1} \over {#2}}}

\def\frac#1#2{{\textstyle{{#1}\over {#2}}}}

\def\lsim{\mathrel{\rlap{\lower4pt\hbox{\hskip1pt$\sim$}}
    \raise1pt\hbox{$<$}}}
\def\gsim{\mathrel{\rlap{\lower4pt\hbox{\hskip1pt$\sim$}}
    \raise1pt\hbox{$>$}}}
\def\sqr#1#2{{\vcenter{\vbox{\hrule height.#2pt
         \hbox{\vrule width.#2pt height#1pt \kern#1pt
         \vrule width.#2pt}
         \hrule height.#2pt}}}}

\def\etal{{\it et al.}}

\def\pt#1{\phantom{#1}}

\def\tt{$t^{\ka\la\mu\nu}$}

\newcommand{\beq}{\begin{equation}}
\newcommand{\eeq}{\end{equation}}
\newcommand{\bea}{\begin{eqnarray}}
\newcommand{\eea}{\end{eqnarray}}
\newcommand{\bit}{\begin{itemize}}
\newcommand{\eit}{\end{itemize}}
\newcommand{\rf}[1]{(\ref{#1})}

\begin{document}

\newcommand{\refeq}[1]{(\ref{#1})}
\def\etal {{\it et al.}}

\title{The SME with gravity and explicit diffeomorphism breaking}

\author{R.\ Bluhm$^1$}

\address{$^1$Department of Physics and Astronomy, Colby College,\\
Waterville ME 04901, USA}

\begin{abstract}
This overview looks at what happens when the Standard-Model 
Extension (SME) is used to investigate gravity theories
with explicit diffeomorphism breaking.
It is shown that when matter-gravity couplings are included, 
the SME generally maintains consistency with the Bianchi identities,
and it therefore provides a useful phenomenological framework for investigating
the effects of explicit diffeomorphism breaking in gravity theories.
\end{abstract}

\bodymatter

\section{Introduction}

In the gravity sector of the SME,
it is traditionally assumed that the SME coefficients arise as
vacuum expectation values in a process of spontaneous diffeomorphism
and local Lorentz breaking.\cite{akgrav04}
With spontaneous breaking,
since all the fields are dynamical,
consistency with geometric identities
such as the Bianchi identities is assured.
Furthermore,
it was spontaneous Lorentz breaking that provided one of the original motivations 
for studying Lorentz breaking,
since it was shown that mechanisms in string field theory 
can lead to this form of spacetime symmetry breaking.\cite{ks}

In addition, when the symmetry breaking is spontaneous,
the excitations of the background fields take
the form of Nambu-Goldstone (NG) 
and massive Higgs-like excitations.
For example, the diffeomorphism NG excitations take the form of Lie derivatives
acting on the vacuum values.
With this known form for the NG modes, 
systematic procedures for developing the post-Newtonian limit
of the SME, including matter-gravity couplings, have been carried out
and have been used to place experimental bounds on 
gravitational interactions that break spacetime symmetry.\cite{qbak06,akjt}

In flat spacetime,
when the SME coefficients are assumed to be constant at leading order,
energy-momentum conservation holds,
and it is not crucial whether the Lorentz violation is spontaneous versus explicit.
Instead, the SME coefficients are viewed simply as fixed backgrounds,
which can be bounded experimentally.
However, when gravity is included in the SME,
and the SME coefficients are treated as nondynamical backgrounds
that explicitly break diffeomorphisms,
then conflicts with the Bianchi identities can arise.  
It is for this reason,
that the spacetime symmetry breaking is traditionally assumed to be spontaneous 
when gravity is included.

Nonetheless, gravity theories with explicit spacetime breaking are of 
interest as modifications of general relativity.
Examples include massive gravity and Horava gravity. 
The question of whether the SME can be used to test for signals of
spacetime symmetry breaking in gravity theories with explicit breaking 
is therefore relevant from both a theoretical and experimental perspective.

\section{SME with Explicit Breaking}

To consider the SME with gravity and explicit spacetime symmetry breaking,
consider an action of the form
\beq
S_{\rm SME} = {\Huge \int} d^4x \sqrt{-g} \, \left[   \fr 1 {2} R +
{\cal L}_{\rm LI} (g_{\mu\nu},f_\si) 
+ {\cal L}_{\rm LV} (g_{\mu\nu},f_\si, \bar k_{\mu\nu\cdots}) \right] .
\label{S}
\eeq
The Lagrangian is separated into an Einstein-Hilbert term
(with $8 \pi G = 1$),
a Lorentz-invariant (LI) term, and a Lorentz-violating (LV) term.
The field $f_\si$ represents a generic dynamical matter field,
while $\bar k_{\mu\nu\cdots}$ generically denotes an SME coefficient
in the context of a theory with explicit breaking.
The SME coefficient in this context is a fixed nondyamical background,
which has no equations of motion.
Its inclusion in the action explicitly breaks diffeomorphism invariance.

If a vierbein formalism is used, components of the background field
with respect to a local Lorentz basis can appear in the Lagrangian as well.
In this case, the background field explicitly breaks local Lorentz invariance 
in addition to diffeomorphisms,
and the consequences of both explicit diffeomorphism breaking and
local Lorentz breaking can be explored in the context of
Riemann-Cartan geometry.\cite{rb15}
However, for the sake of brevity in this overview, only the SME in Riemann
spacetime is considered, and the focus is restricted to explicit breaking of
diffeomorphisms.  

Despite the fact that $S$ is not invariant under diffeomorphisms,
the action must still be invariant under general coordinate transformations
in order to maintain observer independence.
By considering infinitesimal general coordinate transformations,
with $x^\mu \rightarrow x^{\prime \mu}(x) = x^\mu - \xi^\mu$,
four identities follow as a result of $S$ remaining unchanged.\cite{rbas16}
When the matter fields are put on shell, 
and the Bianchi identities are applied,
these identities have the form
\beq
D_\mu T^{\mu\nu} 
- \fr {\de {\cal L}_{\rm LV}} {\de \bar k_{\al\be\cdots}} D^\nu \bar k_{\al\be\cdots} 
+ g^{\nu\si} D_\al (\fr {\de {\cal L}_{\rm LV}} {\de \bar k_{\al\be\cdots}}  \bar k_{\si\be\cdots} ) 
+ \cdots
= 0 ,
\label{identities}
\eeq
where $T^{\mu\nu} $ is the energy-momentum tensor and
$\fr {\de {\cal L}_{\rm LV}} {\de \bar k_{\al\be\cdots}}$ represents the Euler-Lagrange
expression for the background.
In a theory where the background is dynamical,
such as when the symmetry breaking is spontaneous,
the Euler-Lagrange expression $\fr {\de {\cal L}_{\rm LV}} {\de \bar k_{\al\be\cdots}}$ vanishes on shell,
and consistency with covariant energy-momentum conservation follows.
However, with explicit breaking, the background is nondynamical and
its Euler-Lagrange equations need not hold,
which potentially leads to a conflict with the requirement of
covariant energy-momentum conservation.

Evidently, consistency with $D_\mu T^{\mu\nu} = 0$ requires the four equations, 
\beq
- \fr {\de {\cal L}_{\rm LV}} {\de \bar k_{\al\be\cdots}} D^\nu \bar k_{\al\be\cdots} 
+ g^{\nu\si} D_\al (\fr {\de {\cal L}_{\rm LV}} {\de \bar k_{\al\be\cdots}}  \bar k_{\si\be\cdots} ) 
+ \cdots
= 0 ,
\label{identities2}
\eeq
to hold on shell even though $\fr {\de {\cal L}_{\rm LV}} {\de \bar k_{\al\be\cdots}} \ne 0$.
Interestingly,
these equations can in general hold in a theory with explicit breaking,
despite the fact that the Euler-Lagrange equations for $\bar k_{\mu\nu\cdots}$ do not hold.
This is because there are four extra degrees of freedom in the theory due to
the loss of local diffeomorphism invariance.
As long as these four additional degrees of freedom are not suppressed or decouple,
then they can take values that can satisfy \rf{identities2} on shell.

A St\"uckelberg approach can be used to reveal the four
extra degrees of freedom that occur with explicit breaking.
In this approach, four dynamical scalars $\ph^A$ with $A = 0,1,2,3$, 
are introduced through the substitution
$\bar k_{\mu\nu\cdots}(x) \rightarrow D_\mu \ph^A D_\nu \ph^B  \cdots \bar k_{AB\cdots} (\ph) $.
Since the scalars are dynamical,
diffeomorphism invariance is restored.
However, the number of degrees of freedom does not change, 
since the four local symmetries can be used to eliminate the four scalars.
In particular, fixing the scalars as $ \ph^A = \de^A_\mu x^\mu$ gives back
the original explicit-breaking theory in terms of $\bar k_{\mu\nu\cdots}(x)$.

By expanding about the fixed values of the scalars, 
letting $\ph^A = \de^A_\mu (x^\mu + \pi^\mu)$,
the role of the four additional excitations, 
denoted as $\pi^\mu$,
can be examined.
What is found is that the resulting St\"uckelberg excitations take the same
form as in a Lie derivative acting on the background,
\beq
{\cal L}_\xi \bar k_{\mu\nu} \simeq
(D_\mu \pi^\al) \bar k_{\al\nu\cdots}
+ (D_\nu \pi^\al) \bar k_{\mu\al\cdots}
+ \cdots
+ \pi^\al D_\al \bar k_{\mu\nu\cdots} .
\label{NGmodes}
\eeq
What this means is that the four extra degrees of freedom occurring with explicit breaking 
have the same form as the NG modes in a corresponding theory with spontaneous breaking.
Effectively, the St\"uckelberg approach introduces only the four NG modes that
are minimally required to maintain consistency,
while no massive Higgs-like excitations are generated.

In the pure-gravity sector of the SME in a linearized post-Newtonian limit,  
the explicit-breaking background $\bar k_{\mu\nu\cdots}$ couples to
the linearized curvature tensor $R^\ka_{\pt{\ka}\la\mu\nu}$.
If the St\"uckelberg approach is used in this case, 
then the NG modes, $\pi^\mu$, couple to the linearized curvature tensor
and its contractions.
However, the linearized curvature tensor is invariant under diffeomorphisms
at leading order.
This means the NG modes can be gauged away while $R^\ka_{\pt{\ka}\la\mu\nu}$
does not change.
Hence, in the pure-gravity sector,
a useful post-Newtonian limit does not result,
since the NG modes decouple at leading order.\cite{rb15}

However, when matter-gravity couplings are included,
the background SME tensors $\bar k_{\mu\nu\cdots}$ have additional
couplings to matter terms in the action that are not invariant under
diffeomorphisms at leading order.
Thus, in this case,
the NG modes do not decouple,
and the same procedures that were carried out in the 
case of spontaneous breaking\cite{akjt}
can therefore be used as well when the symmetry breaking is explicit.

\section{Conclusion}

In the SME with gravity,
when diffeomorphisms are explicitly broken by couplings to a
nondynamical background SME coefficient $\bar k_{\mu\nu\cdots}$,
the St\"uckelberg approach reveals that the four extra degrees of freedom
that result from the symmetry breaking have the same form as the
four NG modes that appear in a corresponding theory with spontaneous breaking.
As long as these modes are not suppressed or do not decouple,
as is generally the case in a linearized treatment that includes matter-gravity couplings,
then the same procedures can be used with explicit breaking that were developed 
assuming spontaneous breaking.
In particular, the SME can be used to search for potential experimental signals 
that might occur in matter-gravity couplings in gravity theories with explicit
spacetime symmetry breaking.

\end{document}